\documentclass{PoS}
\usepackage{wrapfig}
\usepackage{amsmath}
\usepackage{comment}
\usepackage[left]{lineno}

\newcommand{\hyphen}{\mathchar`-}

\newcommand{\Figref}[1]{Figure.~\ref{#1}}
\newcommand{\BLEFTPol}{e^{-}_{L}e^{+}_{R}}
\newcommand{\BRIGHTPol}{e^{-}_{R}e^{+}_{L}}
\newcommand{\BLEFT}{e^{-},e^{+}}
\newcommand{\BRIGHT}{e^{-},e^{+}}

\title{Sensitivity to anomalous $ZZH$ couplings at the ILC}

\ShortTitle{Sensitivity to anomalous $ZZH$ couplings at the ILC}

\author{\speaker{Tomohisa Ogawa}\\
        The Graduate University for Advanced Studies (SOKENDAI)\\
        E-mail: \email{ogawat@post.kek.jp}}

\author{Keisuke Fujii\\
        High Energy Accelerator Research Organization (KEK)\\
        E-mail: \email{keisuke.fujii@kek.jp}}

\author{Junping Tian\\
        International Center for Elementary Particle Physics (ICEPP/University of Tokyo)\\
        E-mail: \email{tian@icepp.s.u-tokyo.ac.jp}}

\abstract{
\vspace{1mm}
{\it on behalf of the ILD concept group}
\vspace{6mm}
\\
This presentation gives the prospects of measuring the general Lorentz structures of $VVH$ ($V=Z$, $\gamma$ and $W$) couplings at the International Linear Collider (ILC). Sensitivities to Higgs CP-even and CP-odd structures are evaluated by using various Higgs production channels and employing measurements of kinematical distributions. The evaluation is performed based on full detector simulation of the International Large Detector (ILD) at center-of-mass energies $\sqrt{s}=$ 250 and 500 GeV. Combined sensitivities on the anomalous $ZZH$ couplings are provided for a realistic operating scenario of the ILC.}

\FullConference{EPS-HEP 2017, European Physical Society conference on High Energy Physics\\
		5-12 July 2017\\
		Venice, Italy}

\newif\ifPlotOn

\begin{document}

\section{Introduction}

\begin{figure}[htbp]
\vspace{2mm}
\begin{center}
	\begin{tabular}{cc}
	\begin{minipage}{0.375\hsize} 
	\hspace{-11mm}
	\includegraphics[width=71mm]{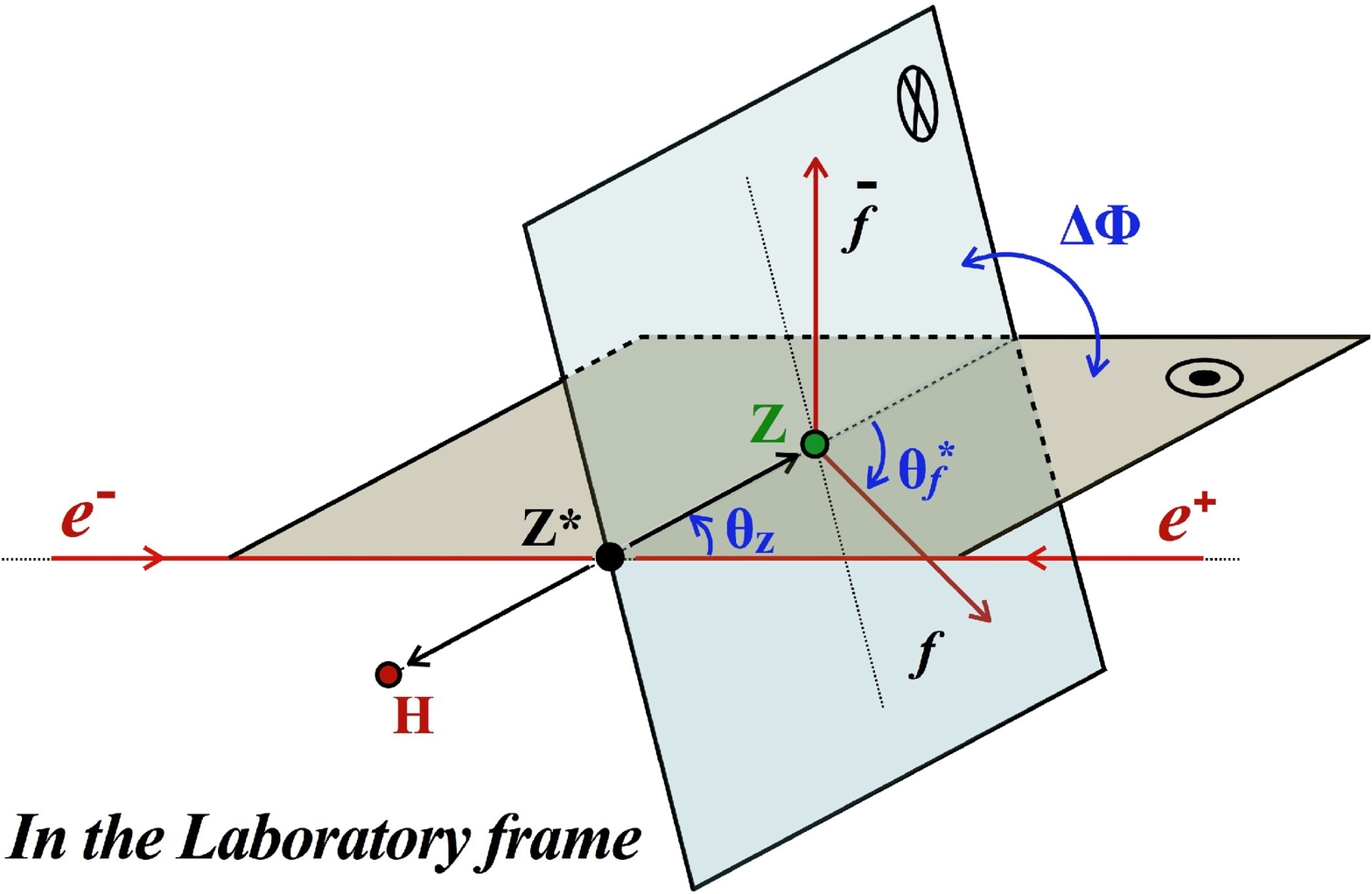}
	\end{minipage}
	\begin{minipage}{0.625\hsize} 
		\begin{tabular}{cc}
		\begin{minipage}{0.46\hsize} 
		\includegraphics[width=48mm]{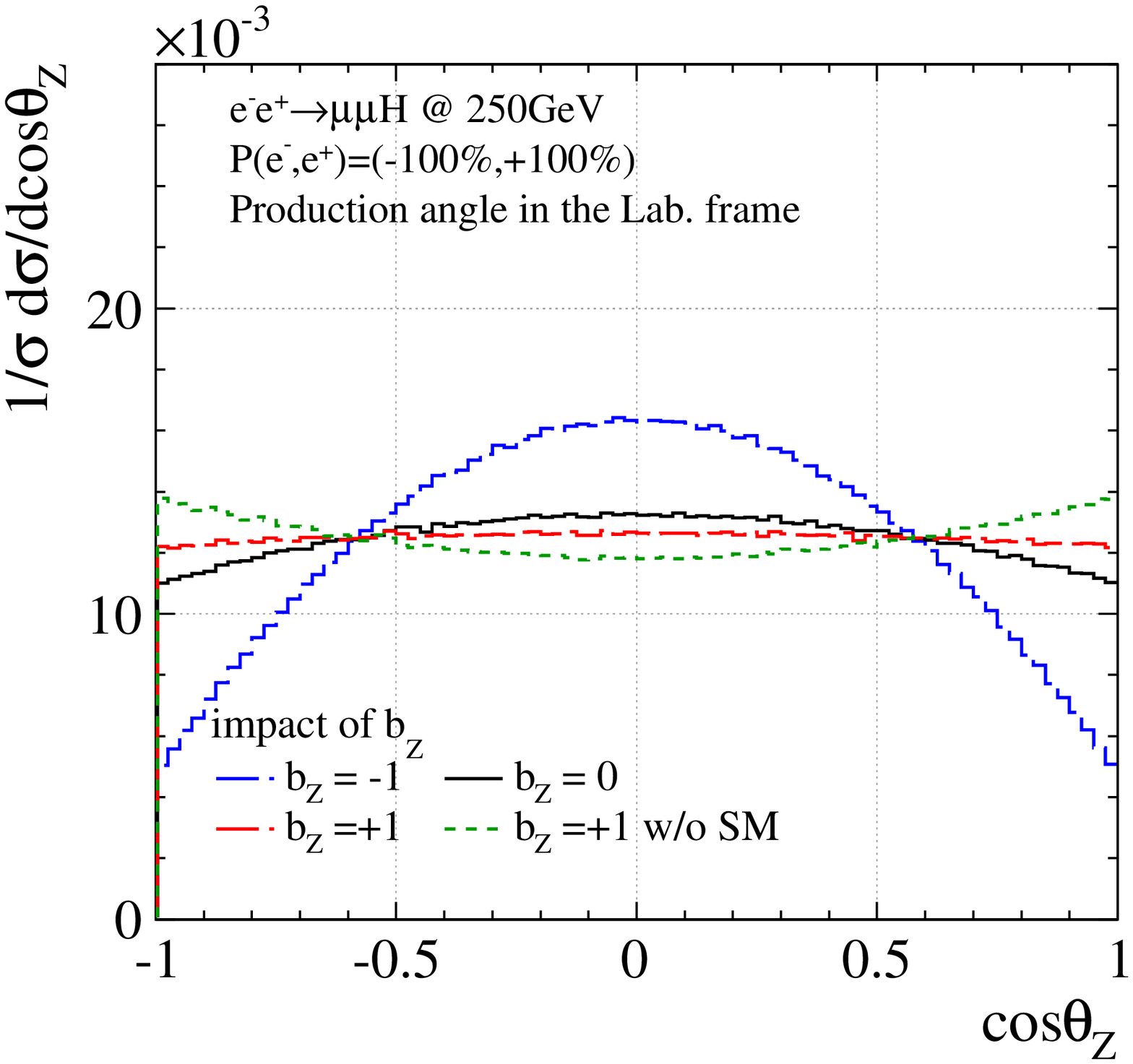}
		\end{minipage}
			\hspace{1mm}
		\begin{minipage}{0.55\hsize} 
		\includegraphics[width=48mm]{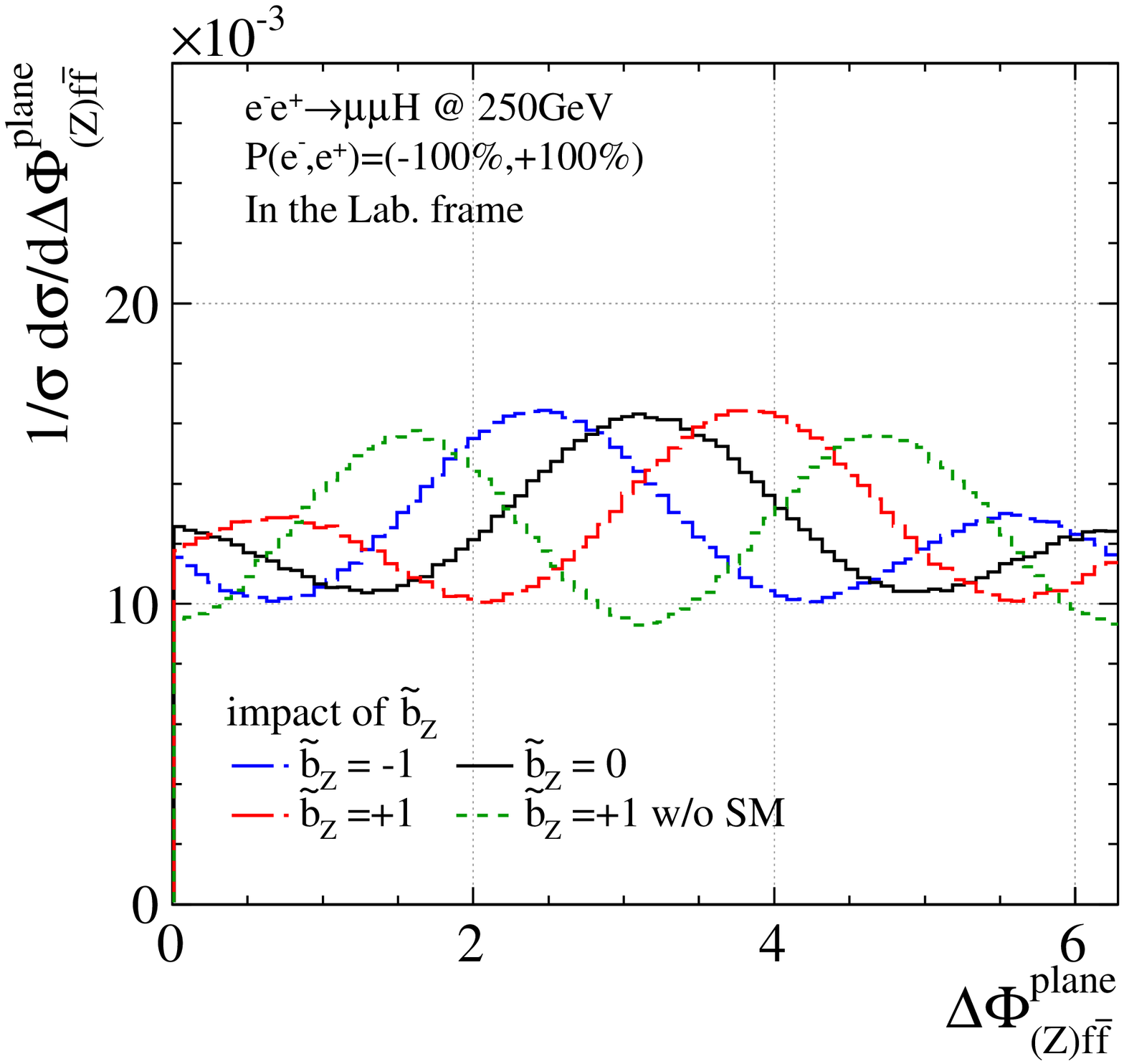}
		\end{minipage}
		\vspace{1mm}
		\\
		\begin{minipage}{0.46\hsize} 
		\includegraphics[width=48mm]{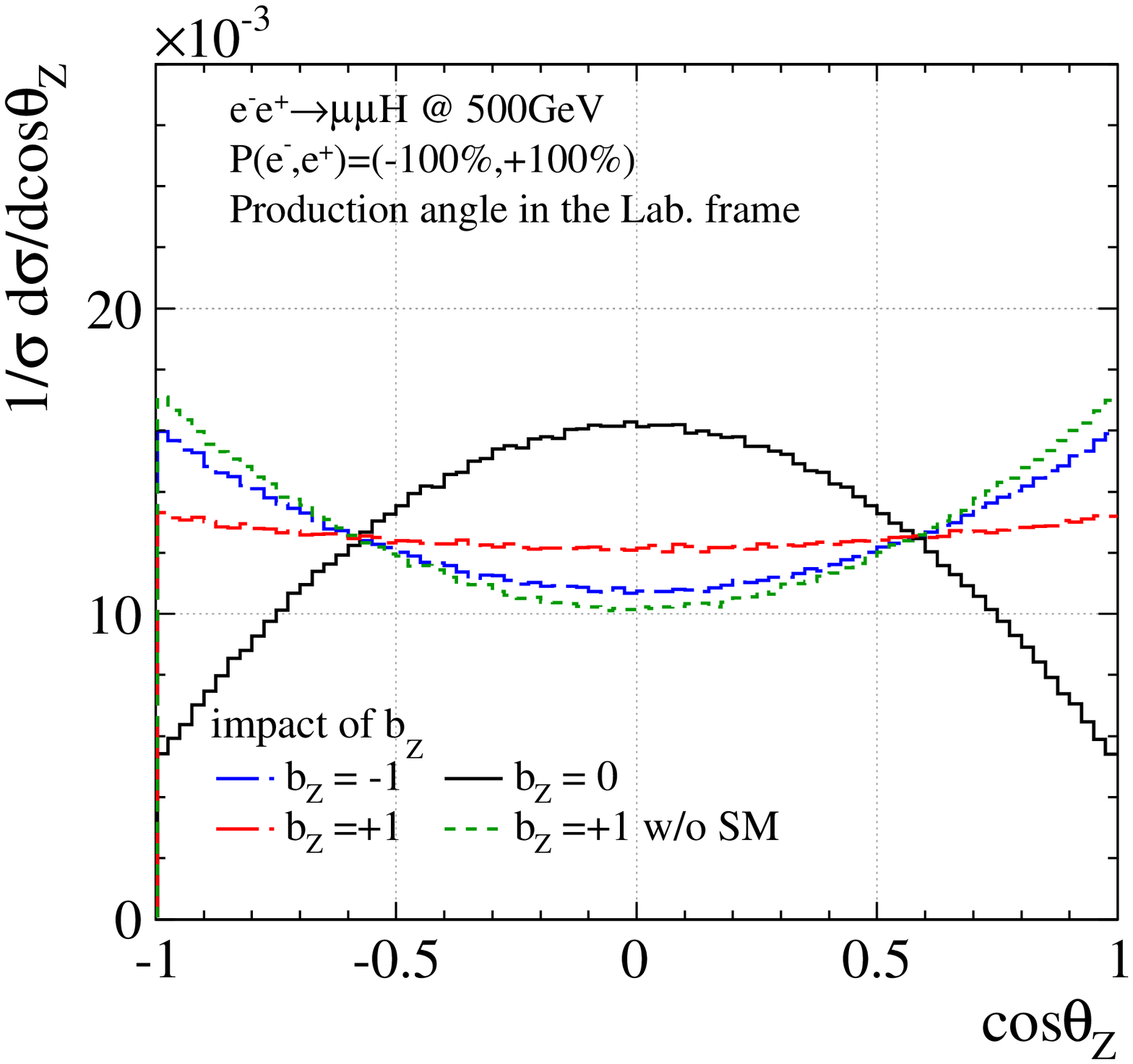}
		\end{minipage}
			\hspace{1mm}
		\begin{minipage}{0.55\hsize} 
		\includegraphics[width=48mm]{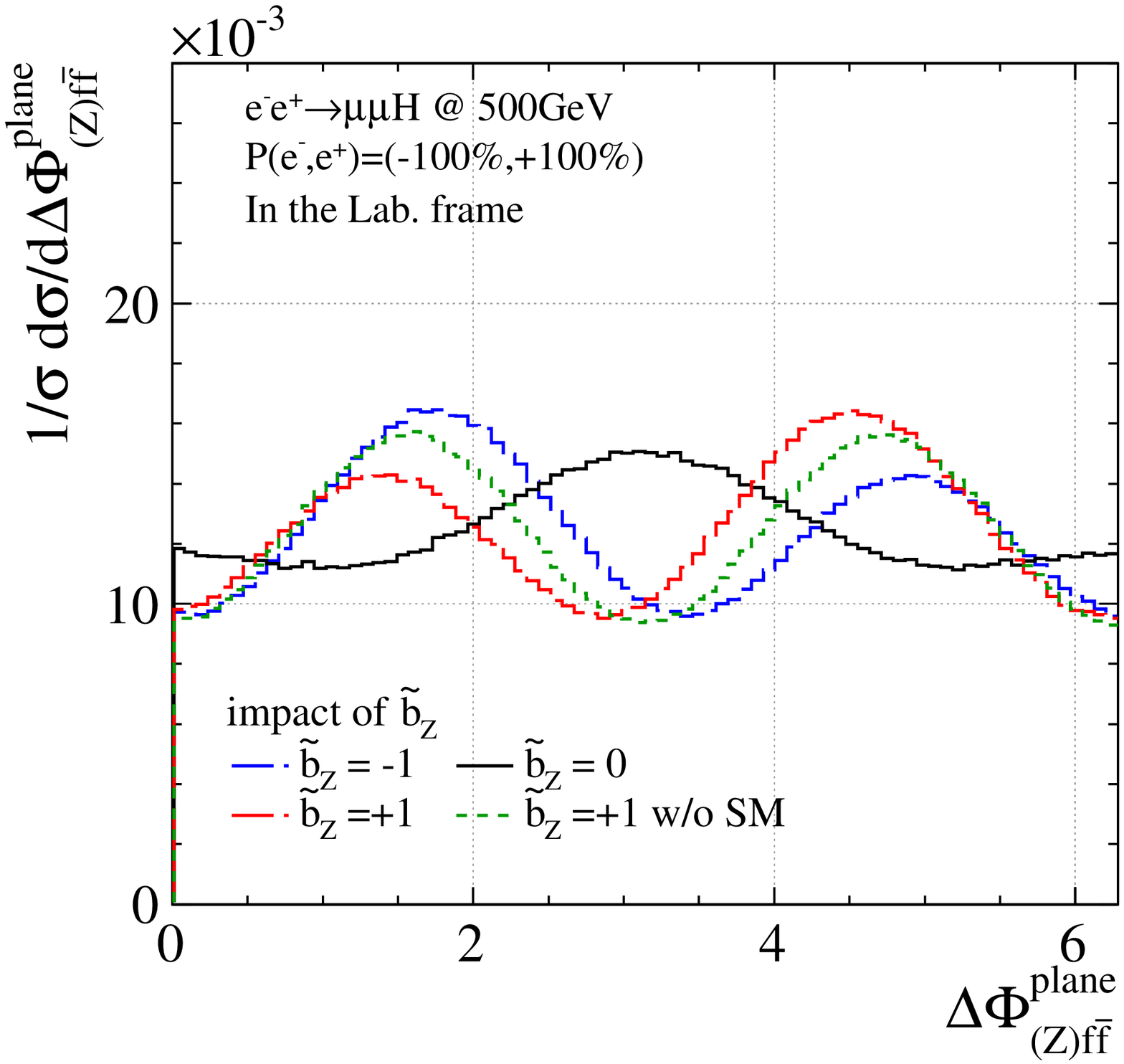}
		\end{minipage}
		\end{tabular}
		\end{minipage}
	\end{tabular}
\caption{(Left) A schematic view of the Higgs-strahlung ($ZH$) process. $\theta_{Z}$ and $\theta_{f}^{*}$ show the production angle of the $Z$ boson and the helicity angle of a daughter fermion of the $Z$ boson. $\Delta\Phi$ shows the angle between two production planes. (Right) Distributions of main observables, $\cos\theta_{Z}$ and $\Delta\Phi$, with different anomalous parameters which are set to be $b_Z=\pm1$ and $\tilde{b}_Z=\pm1$.}
\label{fig:fig3}
		\vspace{-4mm}
\end{center}
\end{figure}

The LHC's discovery of the 125 GeV Higgs boson \cite{ref1,ref2} completed the particle spectrum of the Standard Model (SM) and proved the general idea that the electroweak symmetry was broken by some Higgs field condensed in the vacuum.   
However the SM does not explain why the Higgs condensate formed. 
To reveal this fundamental question new physics describing new phenomena is necessary. And the new physics could manifest itself as anomalies not only in the strengths but also in Lorentz structures of the couplings between the Higgs boson and vector bosons $VVH$ ($V=Z$, $\gamma$ and $W$) \cite{VVH1}. This paper studies the measurement of such anomalies in $ZZH$ couplings, which relate to the electroweak symmetry breaking directly. The study was performed with the simulation framework of the International Large Detector (ILD) concept \cite{ILD} for the International Linear Collider (ILC) \cite{ILC}, and the Lorentz structures of the $ZZH$ couplings are based on an effective Lagrangian from the Effective Field Theory (EFT) \cite{ref4}:
\begin{eqnarray}
\mathcal{L}_{ZZH} = \; M_{Z}^{2}   \Bigr( \frac{1}{{\rm v}} + \frac{a_{Z}}{\Lambda} \Bigr) Z_{\mu}^{}Z^{\mu} H 
				 + \frac{b_{Z}}{2\Lambda} \hat{Z}_{\mu\nu}^{}\hat{Z}^{\mu\nu} H 
				 +  \frac{ \tilde{b}_{Z} }{2\Lambda} \hat{Z}_{\mu\nu}^{}\widetilde{\hat{Z}}^{\mu\nu} H , ~~~~~~~~~~
\label{eq1}
\end{eqnarray}
where ${\rm v}$ is the vacuum expectation value of 246 GeV, $\Lambda$ is the new physics scale which is assumed to be 1~TeV in our study, $a_Z$, $b_Z$ and $\tilde{b}_Z$ are three dimension-less parameters representing anomalous couplings. The field strength tensor $\hat{Z}_{\mu\nu}$ and the dual field strength tensor $\widetilde{\hat{Z}}_{\mu\nu}$ of the $Z$ boson are defined as $\hat{Z}_{\mu\nu} \equiv \partial_{\mu}Z_{\nu} -  \partial_{\nu}Z_{\mu}$ and $\widetilde{\hat{Z}}_{\mu\nu} \equiv \frac{1}{2} \epsilon_{\mu\nu\rho\sigma} \hat{Z}^{\rho\sigma}$. The $a_Z$ term has the same Lorentz structure as that of the SM and affects only the total cross-section while the $b_Z$ and $\tilde{b}_Z$ terms have new tensor structures that can affect angular distributions as well as the total cross-sections for the Higgs-strahlung $e^+e^- \rightarrow ZH$ ($ZH$) and the $ZZ$-fusion $e^+e^-\rightarrow eeH$ ($ZZ$) processes, as illustrated in \Figref{fig:fig3}. 


\section{Analysis on sensitivity to the anomalous $ZZH$ couplings}
\subsection{Monte Carlo samples and a criterion for background suppression}
The Monte Carlo samples used in the study were generated and reconstructed with the framework of the ILD-software for ILC physics and ILD optimization studies \cite{ILD,ref8}. Not only the signal processes but also all SM backgrounds are taken into account for the analysis. Background suppression is performed by maximizing signal significance (=$N_{sig}/\sqrt{N_{sig}+N_{bkg}}$, $N_{sig}$ and $N_{bkg}$ denote the number of events of the signal and background processes), in which a criterion that any angular and corresponding momenta observables are not imposed for the suppression is given. This is because angular information is key observables for determining the anomalous couplings, thus complete insensitive areas become a bias and must be avoided.

%
%
%





\subsection{Strategy for evaluating the anomalous couplings}
Sensitivities to anomalous couplings are evaluated by minimizing $\chi^2$ defined as follows: the first term in the $\chi^2$ formula represents the sensitivity derived from angular information, and the second term from the total cross-section,
\begin{eqnarray}
\!\!\!\!\!\! \chi^{2} = \sum_{i=1}^{n}  \biggl[ \; \frac{   N_{SM} \cdot \frac{1}{\sigma}\frac{d\sigma}{dx} ( x_{i} )  \cdot f_{i}  -  N_{SM} \cdot  \frac{1}{\sigma}\frac{d\sigma}{dx} ( x_{i} ; a_{Z}, b_{Z}, \tilde{b}_{Z}) \cdot f_{i}  \; }{ \Delta n^{obs}_{SM} (x_{i})} \biggr]^{2} + \biggl[    \frac{ N_{SM} - N_{BSM}(a_{Z}, b_{Z}, \tilde{b}_{Z})}{ \delta \sigma \cdot N_{SM} }  \biggr]^{2} , ~~~~  
\label{eq2}
\end{eqnarray}
where $\frac{1}{\sigma}\frac{d\sigma}{dx}( x_{i} )$ and $ \frac{1}{\sigma}\frac{d\sigma}{dx} ( x_{i} ; a_{Z}, b_{Z}, \tilde{b}_{Z})$ are the normalized theoretical angular distributions for the SM and for the non-zero anomalous couplings, respectively, $x$ is an angular variable, $n$ and $i$ denote the total number of bins and bin number in an angular distribution. $f$ is a detector response function for reconstructing the $i$-th bin, which includes detector acceptance, resolution and migration effects. $\Delta n^{obs}_{SM} (x_{i})$ is an observed statistical error for the $i$-th bin. $N_{SM}$ and $N_{BSM}$ are the numbers of expected events for the SM and for non-zero anomalous couplings, respectively. $\delta\sigma$ is the relative total statistical error of the total cross-section and for $ZH$ is taken from the recoil mass studies \cite{refFullSimu2}. 

		

\begin{figure}[tbp]
\begin{center}
\hspace{-2mm}
	\includegraphics[width=153mm]{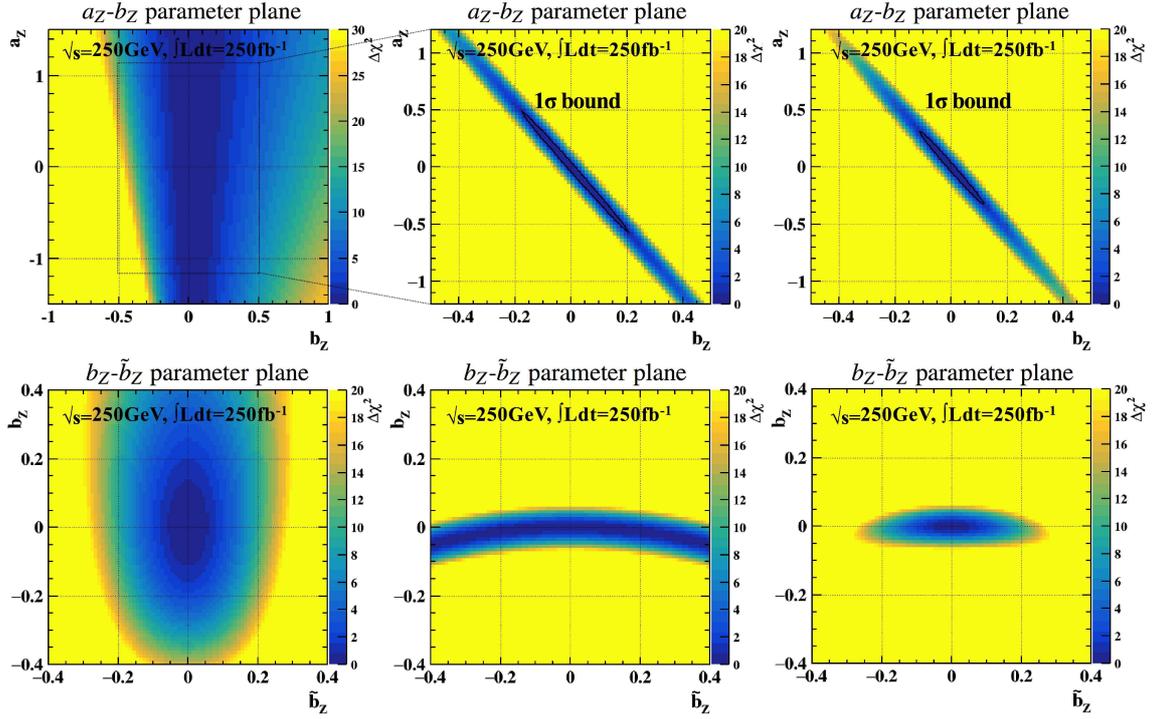}
%
	\vspace{-7mm}
	\caption{$\Delta\chi^2$ distributions in the $a_Z\hyphen b_Z$ parameter plane (upper row) and the $b_Z\hyphen\tilde{b}_Z$ parameter plane (lower row). From left to right, the shape only, the total cross-section only and both information are used for the evaluation of $\Delta \chi^2$. Black lines on the upper middle and right plot correspond to $\Delta\chi^2=1$ bounds. $\sqrt{s}=250$ GeV with the integrated luminosity of $250~{\rm fb}^{-1}$ is assumed. Three channels ($Z\rightarrow\mu\mu$, $ee$ and $ZH\rightarrow qqbb$) of the $ZH$ and one channel ($H\rightarrow bb$) of the $ZZ$ processes are considered as the shape information.}
	\label{fig:fig4}
\end{center}
\end{figure}


\subsection{Effects of shape, total cross-section and $\sqrt{s}$ on the parameter evaluation}
\begin{figure}[tbp]
\begin{center}
\hspace{-2mm}
	\includegraphics[width=153mm]{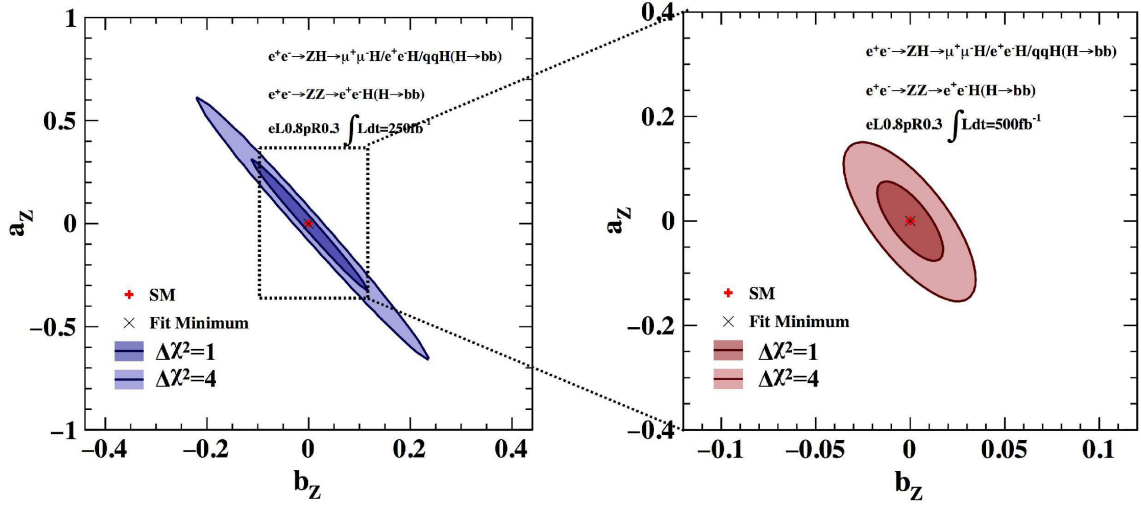}
	\caption{Comparison of the impact of $\sqrt{s}=250$ (left) and $500$ (right) GeV in the $a_Z\hyphen b_Z$ parameter plane. The evaluations are performed with the simultaneous fitting in the three-parameter space and projected onto the $a_Z\hyphen b_Z$ parameter plane, where both of the shape and the cross-section information are used by exploiting four channels of the two processes of the $ZH$ and the $ZZ$ processes. Contours show the bounds of $\Delta\chi^2=1$ and $4$.}
	\label{fig:fig5}
\end{center}
\end{figure}

\Figref{fig:fig4} shows distributions of $\Delta\chi^2$ ($\equiv \chi^2 -\chi^2_{min}$ and $\chi^2_{min}=0$) projected onto the $a_Z\hyphen b_Z$ plane (upper row) and the $b_Z\hyphen\tilde{b}_Z$ plane (lower row) at $\sqrt{s}=250$~GeV. Three plots in each row correspond to evaluations using information on the shape only, the total cross-section only, or both. And both of the $ZH$ and the $ZZ$ processes were included in the evaluations. 
Regarding angular distributions, for three channels ($Z\rightarrow\mu\mu$, $ee$ and $ZH\rightarrow qqbb$) of the $ZH$ process, three-dimensional distributions $x(\cos\theta_Z,\;\cos\theta_f^{*},\;\Delta\Phi)$ are used for both $\sqrt{s}$ = 250 and 500~GeV, and for one channel of the $ZZ$ process ($H\rightarrow bb$) one- and two-dimensional distributions, $x(\Delta\Phi)$ and $x(\cos\theta_h,\;\Delta\Phi)$, are used for $\sqrt{s}$ = 250 and 500~GeV, respectively. 
\textcolor[rgb]{0.0,0.,0}{
The following can be learned from \Figref{fig:fig4}.:
\vspace{-2mm}
\begin{itemize}
 	\item if only the shape information is used, there would be no sensitivity for $a_Z$ when $b_Z$=0. This is because the $a_Z$ term is composed of the same Lorentz structure with that of the SM.
\vspace{ -1mm}
	\item there exists a strong correlation between $a_Z$ and $b_Z$. This is because values of $a_Z$ and $b_Z$ can be adjusted correspondingly to make both of the shape and the total cross-section SM-like. 
\vspace{ 1mm}
	\item the sensitivity of $\tilde{b}_Z$ is almost completely determined by the shape information, and almost uncorrelated with the other two parameters. This is because the CP-odd term has no linear contribution to the total cross-section. 
\end{itemize}
}
\textcolor[rgb]{0.0,0.,0.0}{
\Figref{fig:fig5} plots the 1$\sigma$ ($\Delta\chi^2=1$) and 2$\sigma$ ($\Delta\chi^2=4$) contours in the $a_Z\hyphen b_Z$ plane after performing 3-parameter simultaneous fit under the assumption of $\sqrt{s}=$ 250~GeV with $250~{\rm fb}^{-1}$ and $\sqrt{s}=$ 500~GeV with $500~{\rm fb}^{-1}$ of a beam polarization P$(\BLEFT)= (-80\%,+30\%)$. It clearly shows that at $\sqrt{s}=$500~GeV the correlation between $a_Z$ and $b_Z$ becomes much lower and sensitivities become much better. This is because the $b_Z$ term is momentum dependent.
}

\subsection{Sensitivity to the anomalous $ZZH$ couplings}
The nominal sensitivities showing 1$\sigma$ bounds to the anomalous $ZZH$ couplings and their correlation matrices $\rho$ from 3-parameter simultaneous fit are given as follows, where the integrated luminosities of $250~{\rm fb}^{-1}$ at $\sqrt{s}=250$~GeV and $500~{\rm fb}^{-1}$ at $\sqrt{s}=500$~GeV and two beam polarization states are assumed. For the nominal sensitivities, only the $ZH$ process is used with both shape and total cross-section information.
\vspace{3mm}
\begin{eqnarray}
ZH~~{\rm at}~250~{\rm GeV~with}~~\BLEFTPol: ~~
\left\{ \begin{array}{ l }
	 a_Z =  \pm 0.409 ~~~~~ \\ [0pt]
	b_Z =  \pm 0.147 \\ [0pt]
	\tilde{b}_Z  = \pm 0.066 \\ [0pt]  
\end{array} \!\!\!, ~~ \right.  
   \rho =
  \begin{pmatrix}
   ~~~1~~~ & -0.999 &  ~~~0.006 ~ \\
   \hyphen & 1 & ~~-0.006 \\
   \hyphen & \hyphen & 1
  \end{pmatrix}  ~~~~~ \\
ZH~~{\rm at}~250~{\rm GeV~with}~~\BRIGHTPol:~~
\left\{ \begin{array}{ l }
	 a_Z =  \pm 0.441 ~~~~~ \\ [0pt]
	b_Z =  \pm 0.159\\ [0pt]
	\tilde{b}_Z  = \pm 0.074 \\ [0pt]
\end{array} \!\!\!, ~~ \right.  
   \rho =
  \begin{pmatrix}
   ~~~1~~~ & -0.999~ &  ~-0.006 ~ \\
   \hyphen & 1 & 0.006 \\
   \hyphen & \hyphen & 1
  \end{pmatrix}  ~~~~~ \\
ZH~~{\rm at}~500~{\rm GeV~with}~~\BLEFTPol:~~
\left\{ \begin{array}{ l }
	 a_Z =  \pm 0.123 ~~~~~ \\ [0pt]
	b_Z =  \pm 0.029 \\ [0pt]
	\tilde{b}_Z  = \pm 0.023 \\ [0pt]
\end{array} \!\!\!, ~~ \right.  
   \rho =
  \begin{pmatrix}
   ~~~1~~~ & -0.992~~ &  ~~~ 0.006 ~ \\
   \hyphen & 1 & \!\! -0.009 \\
   \hyphen & \hyphen & 1
  \end{pmatrix}  ~~~~~ \\
ZH~~{\rm at}~500~{\rm GeV~with}~~\BRIGHTPol:~~
\left\{ \begin{array}{ l }
	 a_Z =  \pm 0.132 ~~~~~ \\ [0pt]
	b_Z =  \pm 0.031 \\ [0pt]
	\tilde{b}_Z  = \pm 0.023 \\ [0pt]
\end{array} \!\!\!, ~~ \right.  
   \rho =
  \begin{pmatrix}
   ~~~1~~~ & -0.993 &  ~-0.002 ~ \\
   \hyphen & 1 &~~ 0.001 \\
   \hyphen & \hyphen & 1
  \end{pmatrix}  ~~~~~ 
\end{eqnarray}
%
%
where $\BLEFTPol$ and $\BRIGHTPol$ denote the beam polarization states of P$(\BLEFT)= (-80\%,+30\%)$ and P$(\BRIGHT)=(+80\%,\; -30\%)$, respectively. The combined sensitivities assuming the ILC operating scenario $H20$ \cite{refFullSimu2}, where the total luminosity of 2~${\rm ab}^{-1}$ and 4~${\rm ab}^{-1}$ are accumulated for $\sqrt{s}=$250 and 500~GeV, are given as follows, with both $ZH$ and $ZZ$ processes included.

\begin{eqnarray}
ZH~~{\rm with~the}~~ H20 ~{\rm scenario}:~~
\left\{ \begin{array}{ l }
	 a_Z =  \pm 0.0311 ~~~~~ \\ [0pt]
	b_Z =  \pm 0.0087 \\ [0pt]
	\tilde{b}_Z  = \pm 0.0083 \\ [0pt]
\end{array} \!\!\!, ~~ \right.  
   \rho =
  \begin{pmatrix}
   ~~~1~~~ & -0.911~~ &  ~~~ 0.006 ~ \\
   \hyphen & 1 & \!\! -0.007 \\
   \hyphen & \hyphen & 1
  \end{pmatrix}  ~~~~~ \\
ZH+ZZ~~{\rm with~the}~~ H20 ~{\rm scenario}:~~
\left\{ \begin{array}{ l }
	 a_Z =  \pm 0.0237 ~~~~~ \\ [0pt]
	b_Z =  \pm 0.0070 \\ [0pt]
	\tilde{b}_Z  = \pm 0.0078 \\ [0pt]
\end{array} \!\!\!, ~~ \right.  
   \rho =
  \begin{pmatrix}
   ~~~1~~~ & -0.857 &  ~~~~0.004 ~ \\
   \hyphen & 1 & -0.006 \\
   \hyphen & \hyphen & 1
  \end{pmatrix}  ~~~~~ 
\end{eqnarray}

\vspace{-2mm}
\section{Summary}
Based on the framework of the EFT, the sensitivity to the anomalous $ZZH$ couplings at the ILC was evaluated by exploiting the angular and the total cross-section information using the major Higgs production channels of the Higgs-strahlung and the $ZZ$-fusion processes, for both $\sqrt{s}=250$ and 500~GeV. 
\textcolor[rgb]{0.0,0.,0.0}{Sensitivities in $H20$ scenario to the anomalous couplings $b_Z$ and $\tilde{b}_Z$ with the new Lorentz structures can reach < 1 \%, and the sensitivity to the anomalous coupling $a_Z$ can reach a few \%. This would be very useful to probe the new physics beyond the SM. Once the anomalous $ZZH$ couplings is assumed, anomalous $\gamma ZH$ couplings must be considered since $Z$ and $\gamma$ are mixing with each other through the electroweak symmetry.} However, because of the limit on length, results with the $\gamma ZH$ couplings are not included in this proceedings, but they can be found in the slides of the talk \cite{refX} at the conference EPS-HEP17 and a paper in preparation.
\\
\\
\\
{\small {\bf Acknowledgements:}}  
The authors would like to thank Kiyotomo Kawagoe, Jenny List and Ivanka Bozovic-Jelisavcic for helpful comments on the study. This research was partially supported by JSPS Grants-in-Aid for Science Research No. 16H02176.

\end{document}